\documentclass[12pt]{article}
\usepackage{amsmath}
\def\la{\lambda}

\def\be{\begin{equation}}
\def\ee{\end{equation}}
\def\arr{\begin{array}{rll}}
\def\ea{\end{array}}
\def\bea{\begin{eqnarray}}
\def\eea{\end{eqnarray}}

\def\N2{$N{=}2$}

\def\>{\rangle}
\def\<{\langle}
\def\+{\dagger}
\def\={\ =\ }

\begin{document}
\renewcommand{\thefootnote}{\fnsymbol{footnote}}
\begin{titlepage}
\setcounter{page}{0}
\vskip 1cm
\begin{center}
{\LARGE\bf On dynamical realizations of $l$--conformal}\\
\vskip 0.5cm
{\LARGE\bf  Galilei and Newton--Hooke algebras}\\
\vskip 2cm
$
\textrm{\Large Anton Galajinsky and Ivan Masterov\ }
$
\vskip 0.7cm
{\it
Laboratory of Mathematical Physics, Tomsk Polytechnic University, \\
634050 Tomsk, Lenin Ave. 30, Russian Federation} \\
\vskip 0.4cm
{E-mails: galajin@tpu.ru, masterov@tpu.ru}

\end{center}
\vskip 1cm
\begin{abstract} \noindent
In two recent papers [N. Aizawa, Y. Kimura, J. Segar, J. Phys. A 46 (2013) 405204] and [N. Aizawa, Z. Kuznetsova, F. Toppan, J. Math. Phys. 56 (2015) 031701],
representation theory of the centrally extended $l$--conformal Galilei algebra with half--integer $l$ has been applied so as to construct second order differential equations exhibiting the corresponding group as kinematical symmetry. It was suggested to treat them as the Schr\"odinger equations which involve Hamiltonians describing dynamical systems without higher derivatives. The Hamilto\-nians possess two unusual features, however. First, they involve the standard kinetic term only for one degree of freedom, while the remaining variables provide contributions linear in momenta. This is typical for Ostrogradsky's canonical approach to the descrip\-tion of higher derivative systems. Second, the Hamiltonian in the second paper is not Hermitian in the conventional sense. In this work, we study the classical limit of the quantum Hamiltonians and demonstrate that the first of them is equivalent to the Hamiltonian describing free higher derivative nonrelati\-vistic particles, while the second can be linked to the Pais--Uhlenbeck oscillator whose frequencies form the arithmetic sequence $\omega_k=(2k-1)$, $k=1,\dots,n$. We also confront the higher derivative models with a genuine second order system constructed in our recent work [A. Galajinsky, I. Masterov, Nucl. Phys. B 866 (2013) 212] which is discussed in detail for $l=\frac 32$.
\end{abstract}

\vskip 1cm
\noindent
PACS numbers: 11.30.-j, 02.20.Sv

\vskip 0.5cm

\noindent
Keywords: conformal Galilei symmetry, conformal Newton--Hooke symmetry

\end{titlepage}

\renewcommand{\thefootnote}{\arabic{footnote}}
\setcounter{footnote}0

\noindent
{\bf 1. Introduction}\\

\noindent
Nonrelativistic conformal algebras  \cite{HP,NOR} continue to attract considerable interest owing to the current work on the nonrelativistic AdS/CFT--correspondence.
Conformal extensions of the Galilei and Newton--Hooke algebras are parameterized by a positive half--integer number $l$ such that $(2l+1)$ vector generators
$C^{(n)}_i$, where $i=1,\dots,d$ is a spatial index and $n=0,\dots, 2l$, belong to them \cite{NOR}\footnote{The flat space limit of the $l$--conformal Newton--Hooke algebra in \cite{NOR} does not yield the $l$--conformal Galilei algebra. This shortcoming was overcome in \cite{GM} where the explicit form of admissible central extensions of the $l$--conformal Galilei/Newton--Hooke algebras was established as well.}. $C^{(0)}_i$ and $C^{(1)}_i$ are linked to spatial translations and Galilei boosts while
higher values of $n$ correspond to accelerations.

There are three key issues concerning the $l$--conformal Galilei/Newton--Hooke algebra\footnote{The $l$-conformal Galilei and Newton--Hooke algebras are
isomorphic (see e.g. \cite{NOR,GM}). It is to be remembered, however, that, as far as dynamical realizations are concerned, a linear change of the basis, which links the algebras, implies a change of the Hamiltonian
which alters the dynamics.}. First, dynamical realizations of these algebras constructed so far did not assign any clear physical meaning to the parameter $l$.
Second, apart from the oscillator--like models coupled to external field \cite{FIL}--\cite{GM1}, no interacting theory which exhibits such symmetries is known. Third,
because a number of functionally independent integrals of motion needed to integrate
a differential equation correlates with its order, dynamical realizations of the $l$--conformal Galilei/Newton--Hooke algebra
in general involve higher derivative terms (see, e.g., \cite{LSZ1}--\cite{M} and references therein).

Higher derivative theories typically exhibit instabilities in classical dynamics
and violate unitarity or bring about troubles with ghosts in quantum theory \cite{EW}. An intriguing problem is to understand whether a fully consistent second order interacting system invariant under the $l$--conformal Galilei/Newton--Hooke
group exists such that the acceleration generators are functionally independent.
Note that for the second order models constructed recently in
\cite{FIL}--\cite{GM1} the acceleration generators are redundant. The corresponding integrals of motion can be expressed via those related to spatial translations, Galilei boosts and conformal transformations from $SO(2,1)$ \cite{GM3}.

In two recent works \cite{AKS,AKT}, representation theory of the centrally extended $l$--conformal Galilei algebra with half--integer $l$ has been applied
so as to construct second order differential equations exhibiting the corresponding group as kinematical symmetry.
It was suggested to consider them as the Schr\"odinger equations which involve Hamiltonians describing dynamical systems.
Because the operators are of the second order, it was proposed to treat the resulting models as genuine dynamical systems without higher derivatives.

Two unusual features of the Hamiltonians in \cite{AKS,AKT} ought to be mentioned. First, they involve the standard kinetic term only for one degree of freedom, while the remaining variables provide contributions linear in momenta. Note that this is typical for Ostrogradsky's canonical approach to the description of higher derivative systems (see, e.g., \cite{EW}). Second, the operators in \cite{AKT} are not Hermitian in the conventional sense and a modified scalar product which could render them Hermitian had not been proposed.

In this work, we study the classical limit of the quantum Hamiltonians in \cite{AKS,AKT} and demonstrate that the first of them is equivalent to the Hamiltonian describing free higher derivative nonrelati\-vistic particles, while the second can be linked to the Pais--Uhlenbeck oscillator whose frequencies form the arithmetic sequence $\omega_k=(2k-1) \omega_1$, $k=1,\dots,n$.
As in \cite{AKS,AKT}, our consideration is restricted to half--integer values of $l$ only.
The invariance of the Pais--Uhlenbeck oscillator under the transformations form the $l$--conformal Newton--Hooke group with half--integer $l$
has recently been established in \cite{AGGM}. We also confront the higher derivative models with a genuine second order system \cite{GM3} which is discussed in detail for $l=\frac 32$. In particular,
the symmetry transformations and conserved charges are constructed in explicit form and the redundancy of acceleration generators is demonstrated.

\vspace{0.5cm}

\noindent
{\bf 2. Linking hierarchy of invariant equations to free higher derivative particle }\\

\noindent
In a recent work \cite{AKS}, representation theory of the centrally extended $l$--conformal Galilei algebra with half--integer $l$ in $d=1$ and $d=2$ has been used so as to obtain
a hierarchy of differential equations invariant under the action of the corresponding group. For $d=1$ the first member of the hierarchy reads
\bea\label{lo1}
\left[a_l \mu\left(\frac{\partial}{\partial t}+\sum_{k=1}^{l-\frac{1}{2}}k x_k\frac{\partial}{\partial x_{k-1}}\right)+\frac{\partial^2}{\partial x_{l-\frac{1}{2}}^2}\right]\psi(t,x_i)=0,
\eea
where $a_l=2{\left[\left(l-\frac{1}{2}\right)!\right]}^2$ and $\mu$ is an imaginary mass. It was claimed in \cite{AKS} that (\ref{lo1}) describes a genuine second order system.
Let us demonstrate that (\ref{lo1}) is equivalent to the Schr\"odinger equation for a free nonrelativistic higher derivative particle of the order $2l+1$.

In arbitrary dimension, a free higher derivative particle of the order $2l+1$ is governed by the action functional\footnote{In what follows we omit spatial indices and mark vectors by bold--faced letters.   }
\be\label{S}
S=\frac M2 \int dt {\left(\frac{d^{l+\frac 12} {\bf x}}{d t^{l+\frac 12}} \right)}^2,
\ee
where $M$ is the mass. It is assumed in (\ref{S}) that $l$ is a half--integer number. In Ref. \cite{GK} (see also related works \cite{DH1,AG,AGKM}) this system was shown to exhibit the $l$--conformal Galilei symmetry with half--integer $l$. Quantization of (\ref{S}) based on the Hamiltonian which is built in accord with Ostrogradsky's method leads to the Schr\"odinger equation \cite{GK}
\bea\label{Gom}
\left( i\frac{\partial}{\partial t}+\frac{1}{2M}\frac{\partial^2}{\partial {\bf x}_{l-\frac{1}{2}}^2}+i\sum_{k=1}^{l-\frac{1}{2}} {\bf x}_k\frac{\partial}{\partial {\bf x}_{k-1}}\right) \psi(t,x_i)=0.
\eea
That (\ref{lo1}) is equivalent to (\ref{Gom}) in $d=1$ follows from the redefinition
\bea
\mu=iM,\qquad x_k\,\rightarrow\,\frac{1}{k!}x_k.
\eea

For $d=2$ the first member of the hierarchy of invariant differential equations proposed in \cite{AKS} reads
\bea\label{lo2}
\left[a_l \mu\left(\frac{\partial}{\partial t}+\sum_{k=1}^{l-\frac{1}{2}}k\left(x_k\frac{\partial}{\partial x_{k-1}}+y_k\frac{\partial}{\partial y_{k-1}}\right)\right)+\frac{\partial^2}{\partial x_{l-\frac{1}{2}}y_{l-\frac{1}{2}}}\right]\psi(t,x_i,y_i)=0.
\eea
The simplest way to demonstrate that (\ref{lo2}) describes free higher derivative particles is to rewrite it as the Schr\"odinger equation
\bea
\left(i\frac{\partial}{\partial t}+\frac{1}{2M}\frac{\partial^2}{\partial x_{l-\frac{1}{2}} \partial y_{l-\frac{1}{2}}}+
i\sum_{k=1}^{l-\frac{1}{2}}\left(x_k\frac{\partial}{\partial x_{k-1}}+y_{k}\frac{\partial}{\partial y_{k-1}}\right)\right) \psi(t,x_i,y_i)=0,
\eea
which is obtained from (\ref{lo2}) by the redefinitions
\bea
\mu\,\rightarrow\, iM,\quad x_k\,\rightarrow\,\frac{1}{k!}x_k,\quad y_k\,\rightarrow\,\frac{1}{k!}y_k
\eea
and to focus on the classical limit of the quantum Hamiltonian at hand (for simplicity in what follows we set $M=1$)
\bea\label{h3}
H=\frac{1}{2}p_{x, l-\frac{1}{2}} p_{y, l-\frac{1}{2}}+\sum_{k=1}^{l-\frac{1}{2}}(x_k p_{x, k-1} +y_k p_{y, k-1}).
\eea
Here $p_{x,k}$ and $p_{y,k}$ denote momenta canonically conjugate to $x_k$ and $y_k$, respectively.

The Hamiltonian (\ref{h3}) describes two decoupled higher derivative particles of order $2l+1$ with half--integer $l$. Indeed, for
$l=\frac 32$ Eq. (\ref{h3}) takes the form
\bea
H=\frac{1}{2} p_{x,1} p_{y,1}+x_1 p_{x,0}+y_1 p_{y,0}.
\eea
By applying the linear change of the variables\footnote{Note that (\ref{CT}) is a canonical transformation.}
\bea\label{CT}
\begin{aligned}
&
x_0=\frac{1}{2}(\tilde{x}_0+\tilde{y}_0),\quad \tilde{y}_0=\frac{1}{2}(\tilde{x}_0-\tilde{y}_0),\quad p_{x,0}=(\tilde{p}_{x,0}+\tilde{p}_{y,0}),\quad p_{y,0}=(\tilde{p}_{x,0}-\tilde{p}_{y,0}),
\\[2pt]
&
x_1=\frac{1}{2}(\tilde{x}_1-\tilde{y}_1),\quad y_1=\frac{1}{2}(\tilde{x}_1+\tilde{y}_1),\quad p_{x,1}=(\tilde{p}_{x,1}-\tilde{p}_{y,1}),\quad p_{y,1}=(\tilde{p}_{x,1}+\tilde{p}_{y,1}),
\end{aligned}
\eea
one can bring the Hamiltonian to the form
\bea
H=\left(\frac{1}{2} \tilde{p}_{x,1}^2+\tilde{x}_1 \tilde{p}_{x,0}\right)-\left(\frac{1}{2}\tilde{p}_{y,1}^2+\tilde{y}_1 \tilde{p}_{y,0}\right).
\eea
This is Ostrogradsky's Hamiltonian which describes two decoupled higher derivative particles of the fourth order whose contributions into the full Hamiltonian alternate in sign.

Higher values of half--integer $l$ are treated likewise. For example, for $l=\frac{5}{2}$ the Hamiltonian (\ref{h3}) reads
\bea
H=\frac{1}{2} p_{x,2} p_{y,2}+x_1 p_{x,0}+x_2 p_{x,1}+y_1 p_{y,0}+y_2 p_{y,1},
\eea
which takes the form of the Hamiltonian describing two decoupled higher derivative particles of the sixth order
\bea
H=\left(\frac{1}{2} \tilde{p}_{x,2}^2 +\tilde{x}_2 \tilde{p}_{x,1}+\tilde{x}_1 \tilde{p}_{x,0}\right)-\left(\frac{1}{2}\tilde{p}_{y,2}^2 +\tilde{y}_2 \tilde{p}_{y,1}+\tilde{y}_1 \tilde{p}_{y,0}\right),
\eea
provided the linear canonical change of the variables
\bea
&&
x_0=\frac{1}{2}(\tilde{x}_0-\tilde{y}_0),\quad y_0=\frac{1}{2}(\tilde{x}_0+\tilde{y}_0),\quad p_{x,0}=(\tilde{p}_{x,0}-\tilde{p}_{y,0}),\quad p_{y,0}=(\tilde{p}_{x,0}+\tilde{p}_{y,0}),
\nonumber
\\[2pt]
&&
x_1=\frac{1}{2}(\tilde{x}_1+\tilde{y}_1),\quad y_1=\frac{1}{2}(\tilde{x}_1-\tilde{y}_1),\quad p_{x,1}=(\tilde{p}_{x,1}+\tilde{p}_{y,1}),\quad p_{y,1}=(\tilde{p}_{x,1}-\tilde{p}_{y,1}),
\\[2pt]
&&
x_2=\frac{1}{2}(\tilde{x}_2-\tilde{y}_2),\quad y_2=\frac{1}{2}(\tilde{x}_2+\tilde{y}_2),\quad p_{x,2}=(\tilde{p}_{x,2}-\tilde{p}_{y,2}),\quad p_{y,2}=(\tilde{p}_{x,2}+\tilde{p}_{y,2}),
\nonumber
\eea
has been performed.

We thus conclude that the system (\ref{lo1}) is equivalent to the Schr\"odinger equation for a free nonrelativistic higher derivative particle of the order $2l+1$, while
(\ref{lo2}) describes two decoupled higher derivative particles of the order $2l+1$.

\vspace{0.5cm}

\noindent
{\bf 3. Linking $l$--oscillator to Pais--Uhlenbeck oscillator}\\

\noindent
In a very recent work \cite{AKT}, the so--called $l$--oscillator with $l=\frac 12+\mathbf{N}$ has been introduced which is described by the quantum Hamiltonian
\be\label{lo}
H^{(l)}=-\frac{1}{2m} \partial_{{\bf x}_1}^2+\frac{m}{2} {\bf x}_1^2+\sum_{j=1}^{l-\frac 12} \left((2j+1) {\bf x}_{j+1} \partial_{{\bf x}_{j+1}} -(2l-2j+1){\bf x}_{j} \partial_{{\bf x}_{j+1}}\right)+\frac{(2l-1)(2l+3)}{8}.
\ee
Although a similarity of this system to the Pais--Uhlenbeck oscillator has been observed in \cite{AKT}, it was claimed that the two systems are different as the former is a second order system, while the latter is a higher derivative model.

That the Hamiltonian is a second order differential operator does not mean that the system is free form higher derivatives. The conventional means of quantizing higher derivative models is to construct the Hamiltonian
in accord with Ostrogradsky's prescription (see, e.g., Ref. \cite{EW}). The latter always yields an operator which is at most quadratic in momenta. Higher derivatives of the original classical system manifest themselves in contributions linear in momenta. Note that this is precisely the case for the Hamiltonian (\ref{lo}). Let us demonstrate that the classical limit of (\ref{lo}) can be linked to the Pais--Uhlenbeck oscillator. For simplicity we set $m=1$, $\hbar=1$. As the formulae become increasingly complicated for higher values of half--integer $l$, below we present the analysis for $l=\frac 32$. Further details related to $l=\frac 52$ and $l=\frac 72$ are given in Appendix.

For $l=\frac 32$ the classical limit of (\ref{lo}) reads
\bea\label{32}
H^{(3/2)}=\frac{1}{2} {\bf p}_1^2+3i {\bf x}_2 {\bf p}_2-2i{\bf x}_1 {\bf p}_2+\frac{1}{2} {\bf x}_1^2,
\eea
where $({\bf x}_1,{\bf p}_1)$ and $({\bf x}_2,{\bf p}_2)$ are canonically conjugate pairs obeying the conventional Poisson brackets $\{x_i^\alpha,p_j^\beta\}=\delta_{ij} \delta^{\alpha \beta}$, $\{x_i^\alpha,x_j^\beta\}=0$,
$\{p_i^\alpha,p_j^\beta\}=0$ with $i,j=1,2$ and $\alpha,\beta=1,\dots,d$. Note that the classical partner of (\ref{lo}) turns out to be complex. This means that one should either consider (\ref{lo}) as a physically inconsistent theory or, given the fact that the operator (\ref{lo}) is not Hermitian, allow the classical limit to be complex valued with complex canonical pairs $({\bf x}_1,{\bf p}_1)$ and $({\bf x}_2,{\bf p}_2)$. In this work we choose the second option as subsequent analysis shows that a consistent real dynamics can indeed be associated with the model (\ref{32}).

Deducing the Hamiltonian equations of motion from (\ref{32})
\bea
\dot{{\bf x}}_1={\bf p}_1,\qquad \dot{{\bf p}}_1=-{\bf x}_1+2i{\bf p}_2,\qquad \dot{{\bf x}}_2=3i {\bf x}_2-2i {\bf x}_1,\qquad \dot{{\bf p}}_2=-3i{\bf p}_2,
\eea
one can algebraically express all the variables in terms of ${\bf x}_2$ and its derivatives
\bea\label{2}
{\bf x}_1=\frac{3}{2} {\bf x}_2+\frac{i}{2}\dot{{\bf x}}_2,\qquad {\bf p}_1=\frac{i}{2} {{\bf x}}_2^{(2)}+\frac{3}{2}\dot{{\bf x}}_2,\qquad {\bf p}_2=\frac{1}{4} {{\bf x}}_2^{(3)}-\frac{3i}{4} {{\bf x}}_2^{(2)}+\frac{1}{4}\dot{{\bf x}}_2-\frac{3i}{4}{\bf x}_2,
\eea
where we denoted ${\bf x}_2^{(n)}=\frac{d^n {\bf x}_2}{d t^n}$,
while the equation of motion which governs the dynamics of ${\bf x}_2$ reads
\bea\label{pu32}
&&
{\bf x}_2^{(4)}+10 {{\bf x}}_2^{(2)}+9 {\bf x}_2=0.
\eea
Eq. (\ref{pu32}) describes a complexification of the multi--dimensional Pais--Uhlenbeck oscillator with frequencies of oscillation $\omega_1=1$, $\omega_2=3$ whose invariance under the action of the $l=\frac 32$ conformal Newton--Hooke group has been recently established in \cite{GM1,AGGM}. Because the real and imaginary parts of (\ref{pu32}) describe the same dynamics, at this stage one can consistently truncate the model by considering only the real part of ${\bf x}_2$. This also eliminates an undesirable doubling of states on quantization.

In order to further clarify the connection of (\ref{32}) with the Pais--Uhlenbeck oscillator (\ref{pu32}), let us consider the action functional associated with the latter model
\bea
S=-\frac{1}{8}\int dt\left(\ddot{{\bf x}}_2^2 -10\dot{ {\bf x}}_2^2+9{\bf x}_2^2\right)
\eea
and construct the corresponding Hamiltonian following Ostrogradsky's method. Introdu\-cing Ostrogradsky's canonical variables $({\bf Q}_0,{\bf P}_0)$, $({\bf Q}_1,{\bf P}_1)$
\bea\label{sup}
{\bf Q}_0={\bf x}_2,\qquad {\bf Q}_1=\dot{{\bf x}}_2,\qquad {\bf P}_0=\frac{5}{2}\dot{{\bf x}}_2+\frac{1}{4} {{\bf x}}_2^{(3)},\qquad {\bf P}_1=-\frac{1}{4} {{\bf x}}_2^{(2)},
\eea
and the Hamiltonian
\bea\label{PUH}
H_{PU}^{(3/2)}={\bf Q}_1 {\bf P}_0-2 {\bf P}_1^2-\frac{5}{4} {\bf Q}_1^2+\frac{9}{8} {\bf Q}_0^2,
\eea
one can
invert the relations in (\ref{sup})
\bea
{\bf x}_2={\bf Q}_0,\qquad \dot{{\bf x}}_2={\bf Q}_1,\qquad {{\bf x}}_2^{(2)}=-4 {\bf P}_1,\qquad {{\bf x}}_2^{(3)}=-10 {\bf Q}_1+4 {\bf P}_0,
\eea
and substitute them into the right hand side of (\ref{2}). The result reads
\bea\label{change}
\begin{aligned}
&
{\bf x}_1=\frac{3}{2} {\bf Q}_0+\frac{i}{2} {\bf Q}_1, && {\bf p}_1=\frac{3}{2} {\bf Q}_1-2i {\bf P}_1,
\\[2pt]
&
{\bf x}_2={\bf Q}_0, && {\bf p}_2=-\frac{9}{4} {\bf Q}_1-\frac{3i}{4} {\bf Q}_0+{\bf P}_0+3i {\bf P}_1.
\end{aligned}
\eea
It is then straightforward to verify that the change of the variables (\ref{change}) is canonical. Being substituted into the Hamiltonian (\ref{32}), they yield precisely the Pais--Uhlenbeck oscillator Hamiltonian (\ref{PUH}).

Thus, for $l=\frac 32$ the dynamics associated with the classical limit of the $l$--oscillator proposed in \cite{AKT} can be linked to that of the Pais--Uhlenbeck oscillator with frequencies $\omega_1=1$, $\omega_2=3$.
In a similar fashion one can consider higher values of the half--integer parameter $l$ and demonstrate that the classical limit of (\ref{lo}) can be related to the Pais--Uhlenbeck oscillator whose frequencies form
the arithmetic sequence $\omega_k=(2k-1)$ with $k=1,\dots,n$
\be\label{PPUU}
\prod_{k=1}^{l+\frac 12} \left(\frac{d^2}{d t ^2}+{(2k-1)}^2 \right) {\bf x} (t)=0.
\ee
In particular, the instances of $l=\frac 52$ and $l=\frac 72$ are treated in Appendix. The invariance of (\ref{PPUU}) under the action of the $l$--conformal Newton--Hooke group with $l=\frac 12+\mathbf{N}$
has been established in \cite{AGGM}.

\vspace{0.5cm}

\noindent
{\bf 4. A genuine second order system}\\

\noindent

In a recent work \cite{GM3} (see also \cite{FIL,GM1}), the method of nonlinear realizations was applied to
the $l$--conformal Galilei/Newton--Hooke algebra with the aim to construct a dynamical system
without higher derivative terms in the equations of motion.
A configuration space of the model involves coordinates $\chi_i$, $i=1,\dots,d$, which
parametrize a particle in $d$ spatial dimensions and
a conformal mode $\rho$, which gives rise to an effective external field. The status of the acceleration generators within the scheme was shown to be analogous
to that of the generator of special conformal transformations in $d=1$ conformal
mechanics. Although accelerations are involved in the rigorous
algebraic structure behind the equations of motion,
they prove to be functionally dependent. In \cite{GM3} the general scheme and examples of $l=1$ and $l=2$ were given.
For half--integer $l$ no explicit example, which would include symmetry transformations and conserved charges in explicit form, has been reported.
Below we work out in detail the case of $l=\frac 32$ and confront the results with those in the preceding sections.

According to the analysis in \cite{GM3}, the second order differential equations which hold invariant under the action of the $l=\frac 32$ conformal Galilei group read
\bea\label{sis}
\ddot{\rho}=\frac{\gamma^2}{\rho^3},\qquad \rho^2\frac{d}{dt}\left(\rho^2\frac{d}{dt}\chi_i\right)+\gamma^2\chi_i=0,
\eea
where $\gamma$ is a coupling constant.
The general solution of the equations of motion has the form
\be\label{EM}
\rho(t)=\sqrt{\frac{(\mathcal{D}+t\mathcal{H})^2+\gamma^2}{\mathcal{H}}},\qquad \chi_i(t)=\alpha_i\cos{(\gamma s(t))}+\beta_i\sin{(\gamma s(t))},
\ee
where $\mathcal{D}$, $\mathcal{H}$, $\alpha_i$, $\beta_i$ are constants of integration and the subsidiary function $s(t)$ is given by
\be
s(t)=\frac{1}{\gamma}\arctan{\left(\frac{\mathcal{D}+t\mathcal{H}}{\gamma}\right)}, \qquad \dot s (t)=\frac{1}{\rho(t)^2}.
\ee
The leftmost equation in (\ref{sis}) describes the conventional conformal mechanics in $d=1$, while the particle in $d$ spatial dimensions parametrized by the coordinates $\chi_i$ moves on an ellipse with angular velocity $\frac{d \Phi(t)}{dt}=\frac{\gamma d s(t)}{dt}=\frac{\gamma}{\rho(t)^2}$. Note that the latter is entirely specified by the conformal mode $\rho(t)$ which thus provides a source of an external field.

Following the general scheme in \cite{GM3}, we then construct infinitesimal transformations from the $l=\frac 32$ conformal Galilei group which act on the space of solutions to the equations (\ref{sis})
\bea\label{NewS}
&&
\rho'(t)=\rho(t)+\frac{1}{2}(c+2bt)\rho(t)-(a+bt^2+ct)\dot{\rho}(t),
\nonumber
\\[2pt]
&&
\chi'_i(t)=\chi_i(t)-\left(\frac{\gamma\dot{\rho}}{\rho^2}+\frac{\dot{\rho}^3}{\gamma}\right)\la_i^{(0)}+
\left(\frac{\gamma}{3\rho}+\frac{\rho\dot{\rho}^2}{\gamma}-t\left(\frac{\gamma\dot{\rho}}{\rho^2}+\frac{\dot{\rho}^3}{\gamma}\right)\right)\la_i^{(1)}+
\nonumber
\\[2pt]
&&
+\left(-\frac{\dot{\rho}\rho^2}{\gamma}+2t\left(\frac{\gamma}{3\rho}+\frac{\rho\dot{\rho}^2}{\gamma}\right)-
t^2\left(\frac{\gamma\dot{\rho}}{\rho^2}+\frac{\dot{\rho}^3}{\gamma}\right)\right)\la_i^{(2)}+
\nonumber
\\[2pt]
&&
+\left(\frac{\rho^3}{\gamma}-3t\frac{\rho^2\dot{\rho}}{\gamma}+3t^2\left(\frac{\gamma}{3\rho}+\frac{\rho\dot{\rho}^2}{\gamma}\right)-
t^3\left(\frac{\gamma\dot{\rho}}{\rho^2}+\frac{\dot{\rho}^3}{\gamma}\right)\right)\la_i^{(3)}-(a+bt^2+ct)\dot{\chi}_i(t),
\eea
where $a,b,c,\lambda^{(n)}_i$ are parameters corresponding to time translations, special conformal trans\-formations, dilatations, and vector generators in the algebra, respectively.
It is important to stress that not only does the conformal mode provide a source of an effective external field
for $\chi_i$, but it also enables one to construct transformations corresponding to the vector generators in the algebra, including accelerations.
Considering variations $\delta \rho(t)=\rho'(t)-\rho(t)$, $\delta \chi_i(t)=\chi'_i(t)-\chi(t)$ and computing the
commutator $[\delta_1,\delta_2]$,
one can then reproduce the conventional structure relations of the $l=\frac 32$ conformal Galilei algebra \cite{GM}.

Integrals of motion of the dynamical system (\ref{sis}) corresponding to the infinitesimal symmetry transfor\-ma\-tions displayed above read
\bea
&&
\mathcal{H}=\dot\rho^2+\frac{\gamma^2}{\rho^2}, \qquad \quad  \mathcal{D}=\rho \dot\rho-t \mathcal{H}, \qquad \quad \mathcal{K}=t^2  \mathcal{H}-2t \rho \dot\rho+\rho^2,
\\[2pt]
&&
\mathcal{C}_i^{(0)}=-\rho^2\dot{\chi}_i\left(\frac{\gamma\dot{\rho}}{\rho^2}+\frac{\dot{\rho}^3}{\gamma}\right)+
\chi_i\left(\frac{\gamma^3}{\rho^3}+\frac{\gamma\dot{\rho}^2}{\rho}\right),
\quad
\mathcal{C}_i^{(1)}=\rho^2\dot{\chi}_i\left(\frac{\gamma}{3\rho}+\frac{\rho\dot{\rho}^2}{\gamma}\right)-\frac{2\gamma}{3}\dot{\rho}\chi_i+t\mathcal{C}_i^{0},
\nonumber\\[2pt]
&&
\mathcal{C}_i^{(2)}=-t^2 \mathcal{C}_i^{(0)}+2t\mathcal{C}_i^{(1)}-\frac{1}{\gamma}\dot{\chi}_i\dot{\rho}\rho^4+\frac{1}{3}\gamma\rho\chi_i,
\qquad
\mathcal{C}_i^{(3)}=t^3 C_i^{(0)}-3t^2 C_i^{(1)}+3t C_i^{(2)}+\frac{1}{\gamma}\rho^5\dot{\chi}_i.
\nonumber
\eea
One can verify that constants of the motion $\mathcal{C}_i^{(2)}$ and $\mathcal{C}_i^{(3)}$ which correspond to accelerations are functionally dependent on those related to conformal transforma\-tions, spatial translations and Galilei boosts
\bea
\mathcal{C}_i^{(2)}=\left(\frac{\gamma^2}{3\mathcal{H}^2}-\frac{\mathcal{D}^2}{\mathcal{H}^2}\right)\mathcal{C}_i^{(0)}
-\frac{2\mathcal{D}}{\mathcal{H}}\mathcal{C}_i^{(1)},\qquad
\mathcal{C}_i^{(3)}=\frac{2\mathcal{D}\mathcal{K}}{\mathcal{H}^2}C_i^{(0)}+\frac{3\mathcal{K}}{\mathcal{H}}\mathcal{C}_i^{(1)}.
\eea
This correlates well with the fact that the general solution of the equation of motion for $\chi_i$ can be found from $\mathcal{C}_i^{(0)}$, $\mathcal{C}_i^{(1)}$, $\mathcal{D}$ and $\mathcal{H}$ by purely algebraic means.
Similar redundancy occurs for the generator of special conformal transformation characterizing the $d=1$ conformal mechanics which proves to be functionally dependent on $\mathcal{H}$ and $\mathcal{D}$
\be
\mathcal{K}=\frac{\mathcal{D}^2+\gamma^2}{\mathcal{H}}.
\ee
Note that conformal transformations are essential for the description of the conformal mode $\rho(t)$, while the vector generators
$C^{(n)}_i$ play a key role in the description of $\chi_i(t)$.

We thus conclude that (\ref{sis}) describes a genuine second order system invariant under the action of the $l=\frac 32$ conformal Galilei group in which accelerations generators are redundant.

\vspace{0.5cm}

\noindent
{\bf 5. Conclusion }\\

\noindent
To summarize, in this work we discussed various approaches to the construction of dynamical systems invariant under the $l$--conformal Galilei/Newton--Hooke group with half--integer $l$. In particular, we analyzed
the models advocated in two recent works \cite{AKS,AKT} in the classical limit. It was demonstrated that the first of them was equivalent to free higher derivative nonrelativistic particles of the order $2l+1$, while the second could be linked
to the Pais--Uhlenbeck oscillator whose frequencies form the arithmetic sequence $\omega_k=(2k-1)$, $k=1,\dots,n$.
We suppose that the higher derivative equations of motion in \cite{AKS,AKT} could also be revealed in quantum theory by switching from the Schr\"odinger representation to the Heisenberg picture. It is also likely that the Hamiltonian and positive spectrum attained in \cite{AKT} can be obtained by quantizing the multi--dimensional Pais--Uhlenbeck oscillator of the order $2l+1$ with $l=\frac 12+\mathbf{N}$ whose frequencies form the arithmetic sequence $\omega_k=(2k-1)$ with $k=1,\dots,n$ following the method advocated in \cite{BM}.

A genuine second order system which accommodates the $l=\frac 32$ conformal Galilei symmetry has been proposed. It describes a particle in $d$ spatial dimensions which moves on an ellipse under the influence of an external force caused by an extra conformal mode. As compared to the general scheme in \cite{GM3}, the new results attained in this work include the explicit form of the symmetry transformations and conserved charges. It was also shown that the status of accelerations
is similar to that of the special conformal transformations in $d=1$ conformal mechanics. Although they enter the rigorous algebraic structure behind the equations of motion,
they prove to be functionally dependent. This result correlates well with the order of the differential equations at hand.

The construction of a second order interacting system with positive definite energy which holds invariant under the action of the $l$--conformal Galilei/Newton--Hooke group and in which accelerations are functionally independent remains a challenge.

\vspace{0.5cm}

\noindent{\bf Acknowledgements}\\

\noindent
This work was supported by the MSE program Nauka under the project 3.825.2014/K, the TPU grant LRU.FTI.123.2014, and the RFBR grant 15-52-05022.

\vspace{0.5cm}

\noindent
{\bf Appendix}\\

\noindent
In this Appendix, we display the Hamiltonians of the $l$--oscillator and the Pais--Uhlenbeck oscillator for $l=\frac 52$, $l=\frac 72$ and canonical transformations which link them.

For $l=\frac 52$ the Hamiltonians have the form
\bea\label{3}
&&
H^{(5/2)}=\frac{1}{2}{\bf p}_1^2+3i {\bf x}_2 {\bf p}_2+5i {\bf x}_3 {\bf p}_3-4i {\bf x}_1 {\bf p}_2-2i {\bf x}_2 {\bf p}_3+\frac{1}{2}{\bf x}_1^2,
\nonumber\\[2pt]
&&
H_{PU}^{(5/2)}=32 {\bf P}_2^2+{\bf Q}_2 {\bf P}_1+{\bf Q}_1 {\bf P}_0+\frac{35}{128} {\bf Q}_2^2-\frac{259}{128} {\bf Q}_1^2+\frac{225}{128} {\bf Q}_0^2,
\nonumber
\eea
which are related by the canonical transformation
\begin{align}
&
{\bf x}_1=\frac{15}{8} {\bf Q}_0+i {\bf Q}_1-\frac{1}{8} {\bf Q}_2, && {\bf p}_1=\frac{15}{8} {\bf Q}_1+i {\bf Q}_2-8 {\bf P}_2,
\nonumber\\[2pt]
&
{\bf x}_2=\frac{5}{2} {\bf Q}_0+\frac{i}{2} {\bf Q}_1, && {\bf p}_2=-\frac{15i}{32} {\bf Q}_0+\frac{1}{4} {\bf Q}_1-\frac{49i}{32} {\bf Q}_2-2i {\bf P}_1+16 {\bf P}_2,
\nonumber\\[2pt]
&
{\bf x}_3={\bf Q}_0, && {\bf p}_3=-\frac{45i}{64} {\bf Q}_0-\frac{125}{32} {\bf Q}_1+\frac{125i}{64} {\bf Q}_2+{\bf P}_0+5i {\bf P}_1-25 {\bf P}_2.
\nonumber
\end{align}

Similarly, for $l=\frac 72$ one has the Hamiltonians
\bea
&&
H^{(7/2)}=\frac{1}{2} {\bf p}_1^2+3i {\bf x}_2 {\bf p}_2+5i {\bf x}_3 {\bf p}_3+7i {\bf x}_4 {\bf p}_4- 6i {\bf x}_1 {\bf p}_2-4i {\bf x}_2 {\bf p}_3-2i {\bf x}_3 {\bf p}_4+\frac{1}{2} {\bf x}_1^2,
\nonumber\\[2pt]
&&
H_{PU}^{(7/2)}=-1152 {\bf P}_3^2+{\bf Q}_3 {\bf P}_2+{\bf Q}_2 {\bf P}_1 +{\bf Q}_1 {\bf P}_0+\frac{1225}{512} {\bf Q}_0^2-\frac{3229}{1152} {\bf Q}_1^2+\frac{329}{768} {\bf Q}_2^2-\frac{7}{384} {\bf Q}_3^2,
\nonumber
\eea
which prove to be related by the canonical transformation
\bea
&&
{\bf x}_1=\frac{35}{16} {\bf Q}_0+\frac{71i}{48} {\bf Q}_1-\frac{5}{16} {\bf Q}_2-\frac{i}{48} {\bf Q}_3, ~ {\bf p}_1=\frac{35}{16} {\bf Q}_1+\frac{71 i}{48}{\bf Q}_2-\frac{5}{16} {\bf Q}_3+48i {\bf P}_3,
\nonumber\\[2pt]
&&
{\bf x}_2=\frac{35}{8} {\bf Q}_0+\frac{3i}{2} {\bf Q}_1-\frac{1}{8} {\bf Q}_2, ~ {\bf p}_2=-\frac{35i}{96} {\bf Q}_0+\frac{71}{288} {\bf Q}_1-\frac{5i}{16} {\bf Q}_2+\frac{77}{144} {\bf Q}_3-8 {\bf P}_2-120i {\bf P}_3,
\nonumber\\[2pt]
&&
{\bf x}_3=\frac{7}{2} {\bf Q}_0+\frac{i}{2} {\bf Q}_1, ~ {\bf p}_3=-\frac{35i}{128} {\bf Q}_0+\frac{3}{32} {\bf Q}_1-\frac{2315 i}{1152} {\bf Q}_2-\frac{37}{48} {\bf Q}_3-2i {\bf P}_1+24 {\bf P}_2+218 i {\bf P}_3,
\nonumber\\[2pt]
&&
{\bf x}_4={\bf Q}_0, ~
{\bf p}_4=-\frac{175i}{256} {\bf Q}_0-\frac{12691}{2304} {\bf Q}_1+\frac{12005i}{2304} {\bf Q}_2+\frac{2401}{2304} {\bf Q}_3+{\bf P}_0+7i {\bf P}_1-49 {\bf P}_2-343i {\bf P}_3.
\nonumber
\eea

The action functional corresponding to the Pais--Uhlenbeck oscillator was chosen in the form
\bea
S=-\frac{1}{2 \prod\limits_{k=1}^{l-\frac 12} {(2k)}^2 }\int dt\left({\bf Q}_0   \prod_{k=1}^{l+\frac 12} \left(\frac{d^2}{d t ^2}+{(2k-1)}^2 \right) {\bf Q}_0 \right).
\nonumber
\eea


\begin{thebibliography}{99}
\bibitem{HP}
P. Havas, J. Plebanski, {\it Conformal extensions of the Galilei group and their relation to the Schr\"odinger group}, J. Math. Phys. {\bf 19} (1978) 482.
\bibitem{NOR}
J. Negro, M.A. del Olmo, A. Rodriguez-Marco, {\it Nonrelativistic conformal groups }, J. Math. Phys. {\bf 38} (1997) 3786.
\bibitem{GM}
A. Galajinsky, I. Masterov, {\it Remarks on l-conformal extension of the Newton-Hooke algebra}, Phys. Lett. B {\bf 702} (2011), 265, arXiv:1104.5115.
\bibitem{FIL}
S. Fedoruk, E. Ivanov, J. Lukierski, {\it Galilean conformal mechanics from nonlinear realizations}, Phys. Rev. D {\bf 83} (2011) 085013, arXiv:1101.1658.
\bibitem{GM3}
A. Galajinsky, I. Masterov, {\it Dynamical realization of l-conformal Galilei algebra and oscillators}, Nucl. Phys. B {\bf 866} (2013) 212, arXiv:1208.1403.
\bibitem{GM1}
A. Galajinsky, I. Masterov, {\it Dynamical realizations of l-conformal Newton-Hooke group}, Phys. Lett. B {\bf 723} (2013) 190, arXiv:1303.3419.
\bibitem{LSZ1}
J. Lukierski, P.C. Stichel, W.J. Zakrzewski, {\it Acceleration-extended Galilean symmetries with central charges and their dynamical realizations}, Phys. Lett. B {\bf 650} (2007) 203, hep-th/0702179.
\bibitem{DH1}
C. Duval, P. Horv\'athy, {\it Conformal Galilei groups, Veronese curves, and Newton--Hooke spacetimes}, J. Phys. A {\bf 44} (2011) 335203, arXiv:1104.1502.
\bibitem{GK}
J. Gomis, K. Kamimura, {\it Schrodinger equations for higher order non-relativistic particles and N-Galilean conformal symmetry}, Phys. Rev. D {\bf 85} (2012) 045023, arXiv:1109.3773.
\bibitem{AG}
K. Andrzejewski, J. Gonera, {\it Dynamical interpretation of nonrelativistic conformal groups}, Phys. Lett. B {\bf 721} (2013) 319.
\bibitem{AGKM}
K. Andrzejewski, J. Gonera, P. Kosinski, P. Maslanka, {\it On dynamical realizations of l-conformal Galilei groups}, Nucl. Phys. B {\bf 876} (2013) 309, arXiv:1305.6805.
\bibitem{AGGM}
K. Andrzejewski, A. Galajinsky, J. Gonera, I. Masterov, {\it Conformal Newton-Hooke symmetry of Pais-Uhlenbeck oscillator}, Nucl. Phys. B {\bf 885} (2014) 150, arXiv:1402.1297.
\bibitem{M}
I. Masterov, {\it Higher-derivative mechanics with N=2 l-conformal Galilei supersymmetry}, J. Math. Phys. {\bf 56} (2015) 2, 022902, arXiv:1410.5335.
\bibitem{EW}
D.A. Eliezer, R.P. Woodard {\it The problem of nonlocality in string theory}, Nucl. Phys. B {\bf 325} (1989) 389.
\bibitem{AKS}
N. Aizawa, Y. Kimura, J. Segar, {\it Intertwining operators for l-conformal Galilei algebras and hierarchy of invariant equations}, J. Phys. A {\bf 46} (2013) 405204,
arXiv:1308.0121.
\bibitem{AKT} 	
N. Aizawa, Z. Kuznetsova, F. Toppan, {\it l-oscillators from second-order invariant PDEs of the centrally extended conformal Galilei algebras}, J. Math. Phys. {\bf 56} (2015) 031701, arXiv:1501.00121.
\bibitem{BM}
C.M. Bender, P.D. Mannheim, {\it No--ghost theorem for the fourth--order derivative Pais--Uhlenbeck oscillator model}, Phys. Rev. Lett. {\bf 100} (2008) 110402, arXiv:0706.0207.
\end{thebibliography}
\end{document}